\documentclass[twocolumn,superscriptaddress,notitlepage,%
aps,longbibliography,pra]{revtex4-1}

\usepackage{amsmath}
\usepackage{amssymb}
\usepackage{bm}
\usepackage{bbm}
\usepackage{graphicx}
\usepackage{cancel}
\usepackage{maybemath}
\usepackage{xcolor}
\usepackage{tikz}
\usepackage{tcolorbox}

\newcommand{\ii}{\mathrm{i}}

\newcommand{\calA}{\mathcal{A}}
\newcommand{\calB}{\mathcal{B}}
\newcommand{\calC}{\mathcal{C}}
\newcommand{\calD}{\mathcal{D}}
\newcommand{\calS}{\mathcal{S}}

\newcommand{\scale}{\times 10^{-5}}

\begin{document}

\title{Precision Rydberg State Spectroscopy with Slow Electrons
and Proton Radius Puzzle}

\author{Ulrich D.~Jentschura}
\affiliation{Department of Physics and LAMOR,
Missouri University of Science and Technology,
Rolla, Missouri 65409, USA}

\author{Dylan C. Yost}
\affiliation{Department of Physics, Colorado State University,
Fort Collins, Colorado 80523, USA}

\begin{abstract}
The so-called proton radius puzzle (the current discrepancy
of proton radii determined from spectroscopic 
measurements in ordinary versus muonic hydrogen)
could be addressed via an accurate measurement
of the Rydberg constant, 
because the proton radius and the Rydberg constant values 
are linked through high-precision optical spectroscopy.
We argue that, with manageable additional experimental effort,
it might be possible to improve circular Rydberg
state spectroscopy, potentially leading to an important
contribution to the clarification of the puzzle.
Our proposal involves circular
and near-circular Rydberg states of hydrogen
with principal quantum number around $n=18$,
whose classical velocity on a Bohr orbit
is slower than that of the fastest macroscopic 
man-made object, the Parker Solar Probe.
We obtain improved estimates for the quality factor 
of pertinent transitions, and 
illustrate a few recent improvements in instrumentation
which facilitate pertinent experiments.
\end{abstract}

\maketitle

%
%
\section{Introduction}
\label{sec1}

The Rydberg constant is of consummate importance
for our understanding of fundamental physics.
Notably, this constant is an important input 
datum for the calculation of transition frequencies
in hydrogen and deuterium 
(see Table~II of Ref.~\cite{JeKoLBMoTa2005}
and Refs.~\cite{MoTaNe2008,TiMoNeTa2021}).
In addition to the Rydberg constant,
accurate values of the proton 
and deuteron radii are also required
in order to calculate transition frequencies
in hydrogen and deuterium.
Conversely, one can infer proton and deuteron 
radii from precise values of 
hydrogen and deuterium frequencies
(see Refs.~\cite{JeKoLBMoTa2005,TiMoNeTa2021} and 
Table~45 of Ref.~\cite{MoTaNe2008}).

\begin{figure}[t!]
\begin{center}
\begin{minipage}{0.99\linewidth}
\begin{center}
\includegraphics[width=0.99\linewidth]{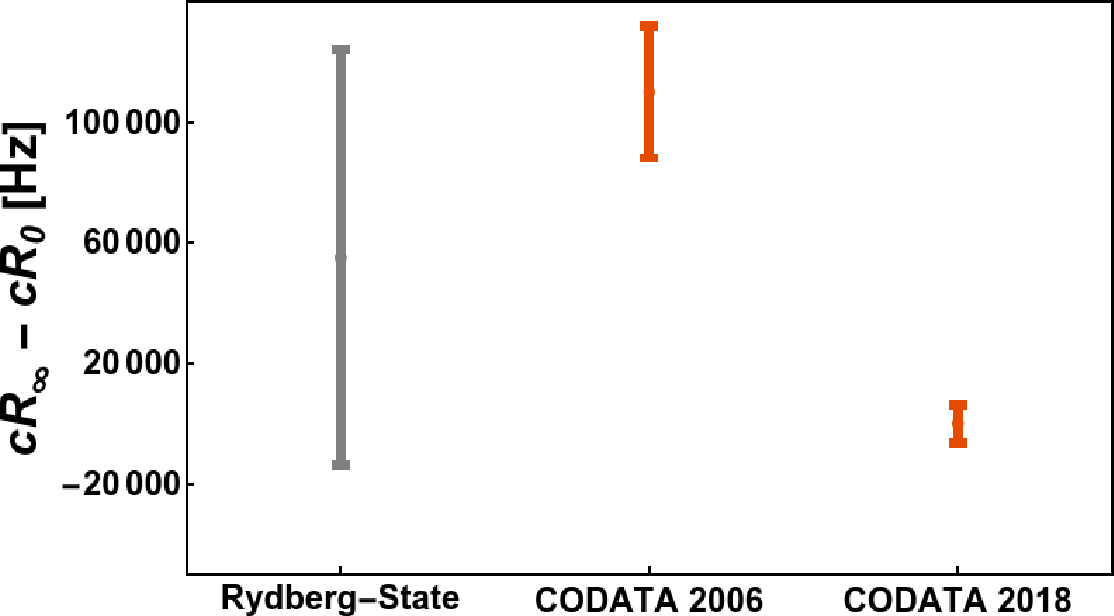}
\caption{\label{fig1} 
We examine the values for the 
Rydberg constant, converted to 
frequency units, from CODATA 
adjustments and from the (unpublished, grayed) 
result communicated in Ref.~\cite{dV2002}.
The CODATA (2006) value was reported in Ref.~\cite{MoTaNe2008},
and the CODATA (2018) value is from
Ref.~\cite{TiMoNeTa2021}.
The reference value $R_0$ is from the 2018 adjustment.}
\end{center}
\end{minipage}
\end{center}
\end{figure}

With the advent of muonic hydrogen
spectroscopic measurements~\cite{PoEtAl2010,PoEtAl2016}, 
the CODATA value of the proton radius 
has shifted from a 2006 value of 
about $R_p \approx 0.88 \, {\rm fm}$ 
to a 2018 value of about
$R_p \approx 0.84 \, {\rm fm}$,
entailing a concomitant change in the 
Rydberg constant~\cite{MoTaNe2008,TiMoNeTa2021}.
From the 2006 to the 2018 
CODATA adjustments~\cite{MoTaNe2008,TiMoNeTa2021},
the Rydberg constant has shifted by much 
more than the uncertainty associated 
with the 2006 value (see Fig.~\ref{fig1}).

One of the most attractive experimental
pathways to the determination of the 
Rydberg constant involves 
highly excited Rydberg states in 
atomic hydrogen, as described in Ref.~\cite{LuEtAl1997}
by a research group working
at the Massachusetts Institute of 
Technology (MIT). Within the same group,
a value for the Rydberg constant was 
obtained in an unpublished thesis
by de Vries~\cite{dV2002} 
(labelled as ``Rydberg state'' in Fig.~\ref{fig1}),
\begin{equation}
\label{cRdeVries}
\left. c R_\infty \right|_{\mathrm{de Vries}} 
= 3\,289\,841\,960\,306(69) \, {\rm kHz} \,,
\end{equation}
which is consistent with the CODATA 2006 value,
and barely consistent with the 2018
CODATA value from Ref.~\cite{TiMoNeTa2021}:
\begin{equation}
\label{cR2018}
\left. c R_\infty \right|_{\rm CODATA,2018} 
= 3\,289\,841\,960\,250(7) \, {\rm kHz} \,.
\end{equation}
The 2006 CODATA value is discrepant,
\begin{equation}
\label{cR2006}
\left. c R_\infty \right|_{\rm CODATA,2006} 
= 3\,289\,841\,960\,360(21) \, {\rm kHz} \,.
\end{equation}
A comparison of the three values 
of the Rydberg constant is made in Fig.~\ref{fig1},
where we use as the reference value 
\begin{equation}
R_0 = \left. R_\infty \right|_{\rm CODATA,2018} \,.
\end{equation}

The situation is interesting because,
before the advent of muonic hydrogen spectroscopy,
values of the Rydberg constant 
and of the proton radius inferred
from  hydrogen and deuterium spectroscopy 
{\em alone} (without any additional input 
from scattering experiments)
were consistent with the 2006 CODATA
values for both the 2006 CODATA value of the 
Rydberg constant, as well as the 
2006 CODATA values of the 
proton and deuteron radii.
This is discussed in detail 
in the discussion surrounding 
Table~45 of Ref.~\cite{MoTaNe2008},
where it is pointed out that the 
proton radius $R_p$, the deuteron radius $R_d$,
and the Rydberg constant can all be 
deduced using input data exclusively 
from hydrogen and deuterium spectroscopy.

Traditionally, the Rydberg constant has been determined on the
basis of Rydberg-state 
spectroscopy of atomic hydrogen~\cite{KiRoShSe1973,Ga1994,%
BiGaJuAl1989,NeEtAl1992,BeEtAl1997,ScEtAl1999,BeEtAl2000review}.
An improved measurement of the 
Rydberg constant would thus constitute 
an important contribution to a resolution
of the proton radius puzzle~\cite{Je2022proton}.
In a remarkable investigation dating about 
20 years back, circular Rydberg states
around quantum numbers $n \approx 30$
have been investigated with the ultimate 
aim of an improved  measurement 
of the Rydberg constant~\cite{dV2002}.
Inspired by the importance of Rydberg states, 
it has been pointed out
in Refs.~\cite{JeMoTaWu2008,JeMoTaWu2009,JeMoTa2010}
that Rydberg-state measurements
in hydrogenlike ions of medium charge numbers
could potentially offer an 
alternative route to a determination
of the Rydberg constant.

The purpose of this paper is threefold.
First, we 
update the calculation of the quality factors 
for transitions among circular
Rydberg states, in comparison to 
the estimate provided in Eq.~(6) of 
Ref.~\cite{JeMoTaWu2008}.
Second, we discuss the 
status of quantum electrodynamic theory 
of Rydberg states, demonstrate
that the theory is very well under 
control on the level of accuracy
required for a determination of the 
Rydberg constant on the level of precision
required for a resolution of the 
proton radius puzzle, and 
discuss the relative suppression 
of a number of notoriously problematic
quantum electrodynamic corrections for 
circular and near-circular Rydberg states.
Calculated values for relativistic Bethe
logarithms for circular 
and near-circular Rydberg states with 
principal quantum numbers $16 \leq n \leq 20$ 
are also provided.
Third, we provide an overview
of recent advances in laser technology 
and other experimental techniques, which facilitate
an improvement of measurements of the
Rydberg constant on the basis 
of Rydberg state measurements.
SI mksA units are employed throughout 
this paper.

%
%
\section{Quality Factors}
\label{sec2}

Of crucial importance for the 
feasibility of high-precision spectroscopy 
experiments are so-called quality factors of 
transitions. The quality factor is the dimensionless
ratio of the transition energy to the 
natural line width of the transition
(measured in radians per second),
where the latter is converted to an energy 
via multiplication by the reduced Planck constant $\hbar$.
Here, we present 
the general formula for the one-photon decay rate 
of a circular Rydberg state, with principal quantum number
$n$ and maximum orbital angular
moment $\ell = n-1$.
This reference state can decay, due to dipole transitions 
to states with principal quantum number
$n-1$ and angular momentum quantum number $\ell = n-2$.
For the decay rate of the 
state with principal quantum number
$n$ and maximum orbital angular
moment $\ell = n-1$, as parameterized by the 
imaginary part of the self energy,
$E = {\rm Re} \, E - \ii \Gamma_n/2$, we find the 
result
\begin{equation}
\Gamma^{\ell=n-1}_n = 
\frac{ 4^{2n} (n-1)^{2n-1} \, n^{2n-4}}{(2n-1)^{4n-1}(2n-3)} \,
\frac{\alpha (Z\alpha)^4 m \, c^2}{3 n^5} \,
\left( \frac{\mu}{m} \right)^3
\end{equation}
which can be expanded for large $n$ as follows,
\begin{multline}
\Gamma^{\ell=n-1}_n = 
\alpha \frac{(Z\alpha)^4 m c^2}{3 n^5} \,
\left( \frac{\mu}{m} \right)^3 \\
\times
\left[ 1 + \frac{3}{2n} + \frac{17}{8 n^2} +
\mathcal{O}\left( \frac{1}{n^3} \right)
\right] \,,
\end{multline}
where $m$ is the electron mass, 
$\mu$ is the reduced mass of the two-body system,
$\alpha$ is the fine-structure constant,
$Z$ is the nuclear charge number, and
the expansion for large $n$ illustrates
that the lifetimes of circular Rydberg states
scale as $n^5$.
The energy difference for transitions among 
circular Rydberg states is 
\begin{equation}
E_n - E_{n-1} = \frac{(Z\alpha)^2 \mu}{2} \, 
\left( \frac{1}{(n-1)^2} - \frac{1}{n^2} \right) \,,
\end{equation}
which scales as $1/n^3$ for large $n$.
Due to the $1/n^5$ asymptotics of the decay 
rate and the $1/n^3$ asymptotics of the 
transition energy, the quality factor increases,
for large $n$,
with the square of the principal quantum number $n$,
\begin{multline}
\label{Q}
Q = \frac{ E_n - E_{n-1} }{ \Gamma^{\ell=n-1}_n + 
\Gamma^{\ell=n-2}_{n-1} } 
\\
= 
\frac{3 n^2}{2 \alpha \, (Z\alpha)^2} \,
\left( \frac{m}{\mu} \right)^2 \,
\left[ 1 - \frac{5}{2 n} - \frac{17}{8 n^2} + 
\mathcal{O}\left( \frac{1}{n^3} \right) \right] \,.
\end{multline}
This formula constitutes an update 
of the estimate given in Eq.~(6) of 
Ref.~\cite{JeMoTaWu2008} (the quality 
factor obtained here is larger by a factor two
as compared to Ref.~\cite{JeMoTaWu2008}).
The estimate in Eq.~\eqref{Q}
illustrates the enormous advantages
of Rydberg states for the measurement of the 
Rydberg constant. The dramatic 
increase of the quality factor 
with the square of the principal quantum number
makes Rydberg state transitions very attractive.
Also, we observe that 
the quality factor is inversely proportional to 
the second power of the nuclear charge number $Z$.
This means that $Z=1$ (atomic hydrogen) offers the 
best quality factor, for given principal quantum 
number $n$.

\begin{figure}[t!]
\begin{center}
\begin{minipage}{0.99\linewidth}
\begin{center}
\includegraphics[width=0.82\linewidth]{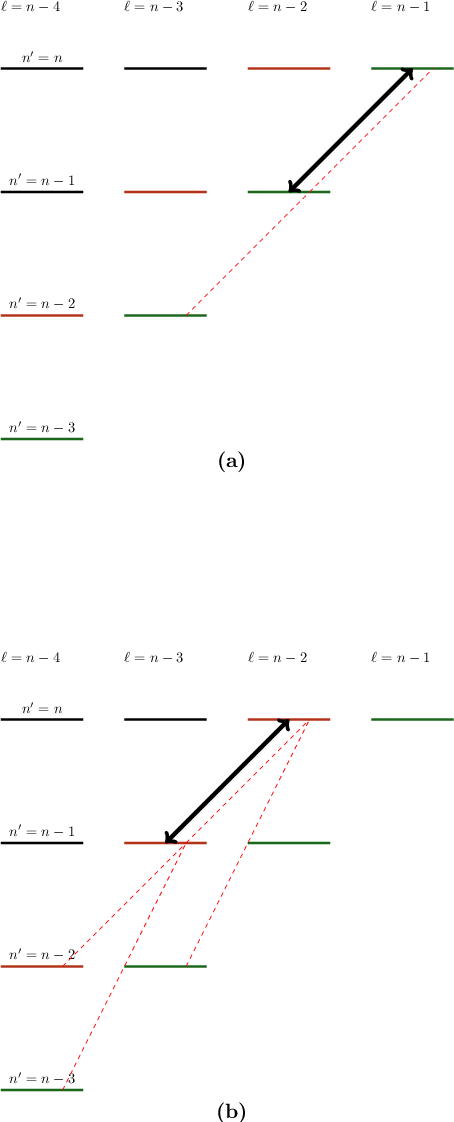}
\caption{\label{fig2}
The level diagram for Rydberg states
illustrates the dipole-allowed transitions
among circular [panel (a)] and
near-circular [panel (b)] states.
Transitions driven for high-precision 
spectroscopy are indicated with two-sided 
arrows. Transitions relevant for the 
calculation of decay rates (quality factors)
are indicated by dashed lines.
Circular Rydberg levels with $\ell = n-1$ 
are marked in green color,
while near-circular Rydberg levels with $\ell = n-2$ 
are marked in red color.}
\end{center}
\end{minipage}
\end{center}
\end{figure}

Let us also evaluate the quality factor for the 
transition among near-circular Rydberg states,
where the upper level has orbital angular 
momentum $\ell = n-2$ and 
the lower level has orbital angular $\ell = n-3$
(see also Fig.~\ref{fig2}).
The calculation of the quality factor 
proceeds in a similar way, 
but one needs to consider two 
available dipole decay channels,
namely, from the reference state with principal quantum 
number $n$ and orbital angular momentum
quantum number $\ell = n-2$,
to lower states with 
$n'=n-1$ and $\ell = n-3$,
and $n'=n-2$ and $\ell = n-3$.
The decay width evaluates to 
\begin{multline}
\label{Gamma2_n}
\Gamma^{\ell=n-2}_n =
\alpha \frac{(Z\alpha)^4 m c^2}{3 n^5} \,
\left( \frac{\mu}{m} \right)^3 
\\
\times \left[ 1 - \frac{1}{2n} - \frac{1}{8 n^2} +
\mathcal{O}\left( \frac{1}{n^3} \right) \right] 
\\
+ \alpha \frac{4 (Z\alpha)^4 m c^2}{3 n^6} \,
\left( \frac{\mu}{m} \right)^3
\\
\times \left[ 1 + \frac{5}{2n} + \frac{25}{4 n^2} +
\mathcal{O}\left( \frac{1}{n^3} \right) \right] \,,
\end{multline}
where the two terms on the right-hand side correspond
to the lower states with $n'=n-1$ and $n'=n-2$, 
respectively. The quality factor evaluates to 
\begin{multline}
\label{Qprime}
Q' = \frac{ E_n - E_{n-1} }{ \Gamma^{\ell=n-2}_n +
\Gamma^{\ell=n-3}_{n-1} }
\\
=
\frac{3 n^2}{2 \alpha \, (Z\alpha)^2}
\left( \frac{m}{\mu} \right)^2 \,
\left[ 1 - \frac{9}{2 n} + \frac{9}{8 n^2} +
\mathcal{O}\left( \frac{1}{n^3} \right) \right] \,,
\end{multline}
which is commensurate with $Q$ given in Eq.~\eqref{Q}
and illustrates that no significant accuracy loss
occurs if one measures near-circular as opposed
to circular Rydberg states.

A quick look at Eqs.~\eqref{cRdeVries},~\eqref{cR2018}
and~\eqref{cR2006}, and Fig.~\ref{fig1},
illustrates that 
one needs to resolve the Rydberg
constant to roughly one part in $10^{11}$ or better
in order to meaningfully distinguish 
between the 2006 and 2018 CODATA values of the 
Rydberg constant.
One can convert this resolution 
to a splitting factor $\calS$, 
which measures the fraction to which 
one needs to split the resonance line in order to 
achieve a resolution of one one part in $10^{11}$.
The splitting factor $\calS$ is given by 
the formula
\begin{equation}
\calS = 10^{11}/Q \,.
\end{equation}
For $Z = 1$, one obtains for $\calS$ the
perfectly reasonable figure
$\calS = 93$ for $n = 18$; expressed differently,
one only needs to split the resonance lines near 
$n=18$ to one part in 93 is order to achieve a resolution
which meaningfully contributes to a resolution
of the proton radius puzzle.

Cross-damping terms (non-resonant corrections)
can be generated by virtual levels displaced
by a fine-structure interval~\cite{JeMo2002}.
A rough estimate of the corresponding 
energy (frequency) shift $\delta E_{\rm CD}$
(we set $\hbar = 1$) is given by the 
expression~\cite{JeMo2002}
\begin{equation}
\label{cross1}
\delta E_{\rm CD} \sim
\frac{\Gamma_n^2}{\delta E} \,,
\end{equation}
Here, $\delta E$ is the displacement
of the virtual state responsible
for the cross-damping energy shift.
As pointed out in Ref.~\cite{JeMo2002},
the nearest virtual states
which can contribute to differential 
cross sections are states displaced
from the upper state of the Rydberg 
transition by a fine-structure interval.
The maximum angular momentum is 
$\ell_{\rm max} = n-1$.
The total angular momenta for the 
circular Rydberg states are
$\ell_{\rm max} \pm 1/2$.
The two possible values for the 
total angular momentum quantum numbers 
of the upper level are thus
$j_+ = n-1/2$ and $j_- = n-3/2$,
one of these being the reference level,
the other being the virtual level
which contributes to the cross damping.
So, we have potential nonresonant 
contributions from virtual levels
with an energy displacement
\begin{equation}
\label{cross2}
\delta E = E_{n,j_+} - E_{n,j_-} =
\frac{ (Z\alpha)^4 m }{ 2 n^4 (n-1)} \approx
\frac{ (Z\alpha)^4 m }{ 2 n^5 } \,.
\end{equation}
The ratio of the cross-damping energy shifts
relative to transition frequency is thus estimated
by the expression
\begin{equation}
\label{cross3}
\chi \equiv 
\frac{\delta E_{\rm CD}}{E_n - E_{n-1}} \sim
\frac29 \frac{\alpha^2 (Z \alpha)^2}{n^2} \,.
\end{equation}
For $Z=1$ and $n=18$, this evaluates 
to $1.9 \times 10^{-12}$, which is less
than the accuracy required in order to 
distinguish between the 2006 and 2018 CODATA
values of the Rydberg constant.
This estimate suggests that cross-damping effects 
are suppressed for Rydberg states
and do not represent an obstacle 
for the determination of the Rydberg 
constant from highly excited, circular 
Rydberg states. 

The above estimates given in 
Eqs.~\eqref{cross1}---\eqref{cross3}
are valid for the differential cross 
section~\cite{JeMo2002}.
For the total cross section,
these estimates improve even further,
consistent with pertinent considerations reported
in Refs.~\cite{Lo1952,JeMo2002,UdEtAl2019}.

%
%
\section{Quantum Electrodynamic Effects}
\label{sec3}

One might ask if the theory of Rydberg-state
transition is well enough under control in order
to facilitate the interpretation of 
a measurement of transitions among Rydberg states.
As outlined in Ref.~\cite{MoTaNe2008},
the theoretical contributions to 
the Lamb shift of Rydberg states,
on the level necessary for a determination
of the Rydberg constant,
can be summarized into just four terms:
{\em (i)} the Dirac energy (in the nonrecoil limit)
which is summarized in Eq.~(1)
of Ref.~\cite{JeMoTaWu2008},
{\em (ii)} the recoil corrections
from the Breit Hamiltonian, which are 
summarized in Eq.~(2)
of Ref.~\cite{JeMoTaWu2008},
{\em (iii)} the relativistic-recoil corrections
summarized in Eq.~(3)
of Ref.~\cite{JeMoTaWu2008},
and {\em (iv)} the self-energy effect
summarized in Eq.~(4)
of Ref.~\cite{JeMoTaWu2008}.
Calculated values of nonrelativistic Bethe logarithms,
which enter the expression for 
the relativistic recoil correction,
have been tabulated for all states with 
principal quantum numbers $n \leq 200$ 
in Ref.~\cite{JeMo2005bethe}.
This favorable situation
illustrates the tremendous simplifications
possible for Rydberg states.
Notably, vacuum-polarization,
nuclear-size, and nuclear-structure corrections
can be completely ignored for circular
Rydberg states whose probability density
at the nucleus vanishes.

Among the four effects listed above,
the most interesting contribution concerns the 
bound-state self-energy $E_{\rm SE}$, which
is described by the formula
\begin{multline}
\label{ESE}
E_{\rm SE} = \frac{\alpha}{\pi}
\frac{(Z\alpha)^4 \, m}{n^3} \,
\biggl( A_{40} + (Z\alpha)^2
\\
\times
\left\{ A_{61} \,
\ln\left[ \frac{m}{\mu} (Z\alpha)^{-2}\right]
+
A_{60} \right\} \biggr) \,.
\end{multline}
The first subscript of the $A$ coefficients
counts the number of $Z\alpha$, 
while the second counts the number of 
logarithms $\ln[ \frac{m}{\mu} (Z\alpha)^{-2}]$.

The general result for the $A_{40}$ coefficient
for circular Rydberg states with orbital angular 
momentum $\ell \neq 0$ and 
principal quantum number $n \geq 2$
is well known,
\begin{equation}
A_{40} = - \left( \frac{\mu}{m} \right)^2 
\frac{1}{2 \kappa (2 \ell + 1)} \, 
- \frac43 \, \left( \frac{\mu}{m} \right)^3 
\ln k_0(n, \ell) \,,
\end{equation}
where $\kappa = (-1)^{j + \ell + 1/2}$ is the Dirac
angular quantum number and 
$\ln k_0(n, \ell)$ is the Bethe logarithm. 
(For values of $\ln k_0(n, \ell)$,
one consults Ref.~\cite{JeMo2005bethe}.) The functional 
dependence on the reduced mass is a consequence
of the proton's convection current; an explanation is 
given in Chap.~12 of Ref.~\cite{JeAd2022book}.
Here, we will place special emphasis on circular,
and near-circular Rydberg states with 
$\ell = n-1$ and $\ell = n-2$, with $n \geq 13$,
and refer to them as the following series 
of states,
\begin{itemize}
\item series $\calA$: $\ell = n-1$, $j = \ell + 1/2$,
$\kappa = -(j+1/2)$,
\item series $\calB$: $\ell = n-1$, $j = \ell - 1/2$,
$\kappa = (j+1/2)$,
\item series $\calC$: $\ell = n-2$, $j = \ell + 1/2$,
$\kappa = -(j+1/2)$,
\item series $\calD$: $\ell = n-2$, $j = \ell - 1/2$,
$\kappa = (j+1/2)$.
\end{itemize}
The $\calA$ series has the highest $\ell$ and $j$ 
for given $n$. The $A_{40}$ coefficients evaluate
to the following expressions for the four series
of states,
\begin{align}
\frac{A_{40}(\calA, n)}{(\mu/m)^2} 
=& \; \frac{1}{2 n(2n-1)} 
- \frac43 \, \frac{\mu}{m}  \, \ln k_0(n, n-1) 
\\
\frac{A_{40}(\calB, n)}{(\mu/m)^2} 
=& \; -\frac{1}{2 (n-1)(2n-1)} 
- \frac43 \, \frac{\mu}{m}  \, \ln k_0(n, n-1) \,,
\\
\frac{A_{40}(\calC, n)}{(\mu/m)^2} 
=& \; \frac{1}{2 (n-1)(2n-3)} 
- \frac43 \, \frac{\mu}{m}  \, \ln k_0(n, n-2) \,,
\\
\frac{A_{40}(\calD, n)}{(\mu/m)^2} 
=& \; -\frac{1}{2 (n-2)(2n-3)} 
- \frac43 \frac{\mu}{m} \ln k_0(n, n-2) \,.
\end{align}
As a function of the principal quantum number,
the Bethe logarithms $\ln k_0(n, n-1)$ and 
$\ln k_0(n, n-2)$ decrease with $n$ for large $n$ 
as $n^{-3}$. In the nonrecoil limit $\mu \to m$,
and the limit of large $n$, one has
\begin{multline}
A_{40}(\calA, n) \approx
-A_{40}(\calB, n) 
\approx A_{40}(\calC, n) 
\\ 
\approx
-A_{40}(\calD, n) \approx \frac{1}{4 n^2} 
\,, \qquad n \to \infty \,.
\end{multline}
The leading quantum electrodynamic corrections
for circular and near-circular Rydberg states
are parameterized by the $A_{40}$ coefficient.
The quantum electrodynamic effects are seen to be suppressed, for large $n$,
by a factor $n^{-2}$ which appears in addition to 
the overall scaling factor $n^{-3}$ in Eq.~\eqref{ESE}.

Higher-loop contributions to the 
anomalous magnetic moment can be taken
into account by the replacement
\begin{equation}
- \left( \frac{\mu}{m} \right)^2
\frac{1}{2 \kappa (2 \ell + 1)} 
\to 
- \left( \frac{\mu}{m} \right)^2
\frac{1}{2 \kappa (2 \ell + 1)} \frac{a_e}{\alpha/(2 \pi)}
\end{equation}
where $a_e$ contains the higher-loop 
contributions to the electron 
anomalous magnetic moment, 
which determines the $g$ factor of 
the electron according to $g = 2(1+a_e)$.
The term $\alpha/(2 \pi)$ is the one-loop
Schwinger value~\cite{Sc1948}.
The quantity $a_e$ can either be taken 
as the most recent experimental value of the 
electron anomalous magnetic moment~\cite{FaEtAl2023},
which results in $a_e = 1.159\,652\,180\,59(13) \times 
10^{-3}$, or as a purely theoretical prediction
including higher-order effects~\cite{La2020}.

The suppression of the quantum electrodynamic
effects for circular and near-circular 
Rydberg states has a physical reason: 
Namely, the velocity of a classical electron
orbiting the nucleus in a Bohr orbit corresponding 
to the principal quantum number $n$ is
\begin{equation}
v_{\rm cl} = \frac{Z \alpha c}{n} \,,
\end{equation}
which evaluates, for $Z=1$ and $n=18$ (this choice of 
$n$ is motivated in Sec.~\ref{sec4}), to
a velocity of $1.21 \times 10^5 {\rm m}/{\rm s}$.
This is slower than the velocity of the 
fastest macroscopic man-made object,
namely, the Parker Solar Probe which recently reached
a velocity of $1.48\times 10^5 {\rm m}/{\rm s}$
on its orbit around the Sun~\cite{PARKER,RaEtAl2023}.
Effects originating from relativity and 
quantum electrodynamics are thus highly suppressed
for circular Rydberg states.

The general result for the $A_{61}$ coefficient,
valid for Rydberg states with $n \geq 13$ and $\ell = n-1$
and $\ell = n-2$, has been given in 
Eq.~(6) of Ref.~\cite{WuJe2008} and 
Eq.~(4) of Ref.~\cite{JeMoTaWu2008} and reads
\begin{equation}
A_{61} = \left( \frac{\mu}{m} \right)^3 \,
\frac{ 3 n^2 - \ell ( \ell + 1 ) }%
{3 n^2 (\ell + 3/2) (\ell + 1) (\ell + 1/2) \ell (\ell - 1/2)} \,,
\end{equation}
a result which is independent of the spin orientation.
This expression evaluates to 
\begin{align}
\frac{A_{61}(\calA, n)}{(\mu/m)^3} 
=& \; \frac{A_{61}(\calB, n)}{(\mu/m)^3} 
= \frac{8}{3 n^2 (n-1) (2n-1) (2n-3)} \,, 
\\
\frac{A_{61}(\calC, n)}{(\mu/m)^3} 
=& \; \frac{A_{61}(\calD, n)}{(\mu/m)^3} 
= \frac{32 (n+2)}{3 n^2 \prod_{i=2}^5 (2n-i) } \,.
\end{align}
In the large-$n$ limit, one has
\begin{multline}
A_{61}(\calA, n)
\approx A_{61}(\calB, n) \\
\approx A_{61}(\calC, n)
\approx A_{61}(\calD, n)
\approx \frac{2}{3 n^5} 
\,, \qquad n \to \infty \,.
\end{multline}
The suppression with $n^{-5}$, in addition to the 
overall scaling factor $n^{-3}$ from Eq.~\eqref{ESE},
again illustrates the smallness
of relativistic and quantum electrodynamic effects
for circular Rydberg states.

\begin{table}[t!]
\caption{\label{table1}
Calculated values for the 
$A_{60}$ coefficients for highly excited 
Rydberg states are given
for the $\calA$, $\calB$, $\calC$ and $\calD$
series of states, 
for principal quantum numbers $16 \leq n \leq 20$.}
\renewcommand{\arraystretch}{1.5}
\begin{center}
\begin{minipage}{1.00\linewidth}
\begin{tabular}%
{c@{\hspace{0.3cm}}c@{\hspace{0.3cm}}c@{\hspace{0.3cm}}c%
@{\hspace{0.3cm}}c@{\hspace{0.3cm}}c}
\hline
\hline
 & & \multicolumn{2}{c}{$\calA$ Series} & \multicolumn{2}{c}{$\calB$ Series} \\
$n$ & $\ell$ & $j$   & $A_{60}(n\ell_j)$      & $j$   &  $A_{60}(n\ell_j)$ \\
\hline
16  & 15     & $\frac{31}{2}$ & $1.059\,675(5) \scale$ & 
               $\frac{29}{2}$ & $0.121\,748(5) \scale$ \\
17  & 16     & $\frac{33}{2}$ & $0.805\,212(5) \scale$ & 
               $\frac{31}{2}$ & $0.078\,287(5) \scale$ \\
18  & 17     & $\frac{35}{2}$ & $0.621\,952(5) \scale$ & 
               $\frac{33}{2}$ & $0.049\,885(5) \scale$ \\
19  & 18     & $\frac{37}{2}$ & $0.487\,434(5) \scale$ & 
               $\frac{35}{2}$ & $0.031\,113(5) \scale$ \\
20  & 19     & $\frac{39}{2}$ & $0.387\,025(5) \scale$ & 
               $\frac{37}{2}$ &  $0.018\,584(5) \scale$ \\
\hline
\hline
 & & \multicolumn{2}{c}{$\calC$ Series} & \multicolumn{2}{c}{$\calD$ Series} \\
$n$ & $\ell$ & $2j$ & $A_{60}(n\ell_j)$      & $2j$ &  $A_{60}(n\ell_j)$ \\
\hline
16  & 14     & $\frac{29}{2}$ & $1.540\,182(5) \scale$ & 
               $\frac{27}{2}$ & $0.155\,784(5) \scale$ \\
17  & 15     & $\frac{31}{2}$ & $1.145\,325(5) \scale$ & 
               $\frac{29}{2}$ & $0.096\,026(5) \scale$ \\
18  & 16     & $\frac{33}{2}$ & $0.867\,820(5) \scale$ & 
               $\frac{31}{2}$ & $0.058\,328(5) \scale$ \\
19  & 17     & $\frac{35}{2}$ & $0.668\,553(5) \scale$ & 
               $\frac{33}{2}$ & $0.034\,217(5) \scale$ \\
20  & 18     & $\frac{37}{2}$ & $0.522\,676(5) \scale$ & 
               $\frac{35}{2}$ & $0.018\,690(5) \scale$ \\
\hline
\hline
\end{tabular}
\end{minipage}
\end{center}
\end{table}

The next-higher coefficient is $A_{60}$, 
which is called the relativistic Bethe 
logarithm~\cite{JeEtAl2003,LBEtAl2003}.
Its absolute magnitude is highly suppressed for 
circular Rydberg states. Specifically,
according to Refs.~\cite{JeMoTaWu2008,JeMoTaWu2009,JeMoTa2010}
and Table 7.2 of Ref.~\cite{Wu2011},
one has 
\begin{multline}
\label{hurrah}
\max\{ | A_{60}(\calA, n) |,
| A_{60}(\calB, n) |, \\
| A_{60}(\calC, n) |,
| A_{60}(\calD, n) | \} 
< 10^{-4} \,,
\qquad n > 13 \,.
\end{multline}
Furthermore, according to the calculations reported
in Ref.~\cite{WuJe2010,JeMoTa2010}, the approximation
$G_{\rm SE} \approx A_{60}$ 
for the nonperturbative 
self-energy remainder function
remains valid to 
excellent approximation for circular Rydberg
states, for low and medium nuclear charge numbers 
(see Table~1 of Ref.~\cite{WuJe2010}
and Tables 1~and~2 of Ref.~\cite{JeMoTa2010}).
The relation~\eqref{hurrah} implies that the correction 
to the transition frequency among circular 
Rydberg states induced by the 
relativistic Bethe logarithm $A_{60}$,
for $Z=1$, is smaller than one part in $10^{-15}$
for $n \geq 13$.
Nevertheless, it is useful to calculate
numerical values of relativistic Bethe
for the states under investigation
here (see Table~\ref{table1}). 
We follow the calculational procedure outlined 
in Ref.~\cite{WuJe2008}.
For calculated values of $A_{60}$ for circular and 
near-circular Rydberg states with $13 \leq n \leq 16$,
we refer to Table~1 of Ref.~\cite{JeMoTaWu2008}
and Table~1 Ref.~\cite{JeMoTa2010}.

%
%
\section{Experimental Considerations}
\label{sec4}

Let us also include a few considerations relevant to the experimental
realization of a high-precision measurement of the Rydberg constant based on
circular Rydberg states.  One might assume that the ultimate experimental
success could be bolstered by choosing transitions with as high a quality
factor $Q$ as possible. As discussed around Eq.~\eqref{Q}, since $Q \propto
n^2$, high $n$ is desirable. 

However, it is also important to consider the sensitivity of a given
measurement to systematic effects.  Many systematic
effects increase with powers of $n$.  For instance, 
shifts and distortions of resonances due to the Stark effect scale as $n^5$~\cite{Lu1988,Ho1998,dV2002},
which produces challenges to measuring
transitions between circular Rydberg states with very high $n$.  However, the previous
measurement between circular Rydberg states of hydrogen
(Ref.~\cite{dV2002}) between $n=27$ and $n=28$, and  $n=29$ and $n=30$, had
negligible contributions from uncertainties in the Stark 
shifts~\cite{dV2002}). The experimental accuracy was instead
limited by dipole-dipole interactions.
Since the dipole moment for an atom in a superposition of adjacent circular
Rydberg states scales as $n^2$, and the systematic effect is related to the
interaction energy of two dipoles, this effect scaled as $n^4$.

Therefore, in order to mitigate the dipole-dipole interactions, it may be
interesting to consider transitions between circular Rydberg states with
somewhat lower $n$.  For instance, with all other experimental parameters being
similar, a transition between $n=18$ and $n=19$ would reduce the
effects of the dipole-dipole interactions by a factor of $\sim 6$ as compared
to the previous measurement~\cite{dV2002}, Another experimental benefit to reducing
$n$ below that demonstrated in \cite{dV2002}, is that blackbody-radiation-induced transitions 
would be mitigated because the thermal
radiation spectral density for temperatures $\leq 300$ K is reduced for the
more energetic transitions occurring between lower-lying states. This may allow the
experiment to be performed at liquid nitrogen as opposed to liquid helium
temperatures.  

The MIT measurement used pulsed lasers at a repetition rate of
$61\,\mathrm{Hz}$ to produce circular Rydberg states.  Therefore, another
option to mitigate dipole-dipole interactions could be to produce a
near-continuous source of circular Rydberg states using 
continuous-wave (cw) lasers.  Since the
dipole-dipole interaction is related to the peak density of circular Rydberg
states, a near-continuous source of circular Rydberg states could allow for a
large reduction in the peak density while maintaining sufficient statistics.
This could be accomplished by first using the $1S$--$2S$ two-photon transition to
populate the $2S$ metastable state as in Refs.~\cite{BeEtAl2017,BrEtAl2022},
followed by excitation to Rydberg levels using a 365\,nm cw laser.  Then
circularization would be performed using the methods outlined in
Ref.~\cite{LuEtAl1997}.

To perform spectroscopy of the $n=18$ to $n=19$ circular Rydberg states, a
millimeter-wave Ramsey apparatus akin to the one employed in
Ref.~\cite{dV2002} could be used.  To excite the transition, a radiation source
at 1.04 THz is needed.  While the millimeter wave source in~\cite{dV2002}
operated at 256 or 316 GHz, a similar source operating at frequencies above 1
THz is possible using a planar GaAs Schottky diode frequency
multiplier~\cite{MeSiLeSc2015}.
The output power of such THz sources is relatively
low.  However, due to the large transition matrix
element between circular Rydberg states, the transition can be saturated with
$<1$ nW and a 3\,mm beam waist.  Therefore, commercially available THz sources
would likely be sufficient~\cite{vadiodes}.

%
%
\section{Conclusions}
\label{sec5}

The main conclusions of this paper are
as follows. In Sec.~\ref{sec2},
we have shown that the quality factors
of transitions among circular Rydberg
are sufficient to comfortably allow for 
a distinction between the 2006 and 2018 
CODATA values of the Rydberg constant
[see Eqs.~\eqref{cR2018} and~\eqref{cR2006},
and Refs.~\cite{MoTaNe2008,TiMoNeTa2021}].
Furthermore, according to the 
considerations reported in Sec.~\ref{sec2},
cross-damping terms do not to present an obstacle to 
such a measurement.
In Sec.~\ref{sec3}, we showed that 
the theory of bound states is sufficiently 
under control to allow for 
a determination of the Rydberg constant
from transitions among circular 
Rydberg states in atomic hydrogen.
Experimental considerations (Sec.~\ref{sec4})
corroborate the advances in technology
which make such a measurement
more feasible than reported in Ref.~\cite{dV2002},
in part by reducing several systematic
effects through a less dense atomic beam which 
can be realized in a continuous-wave 
excitation scheme into the circular 
states.

A few concluding remarks on the proton radius puzzle
are in order. We recall that the proton radius 
puzzle refers to the difference
between the ``smaller'' proton radius of
$R_p \approx 0.84 \, {\rm fm}$
obtained in Ref.~\cite{PoEtAl2010}
and the larger value of 
$R_p \approx 0.88 \, {\rm fm}$
from the 2006 CODATA 
adjustment (see Refs.~\cite{BeEtAl1997,ScEtAl1999,JeKoLBMoTa2005,MoTaNe2008}
and references therein).
Various recent scattering experiments~\cite{BeEtAl2010,XiEtAl2019}
and spectroscopic 
experiments~\cite{BeEtAl2017,FlEtAl2018,BeEtAl2019,BrEtAl2022,YzEtAl2023}
come to conflicting conclusions on the proton radius.
A recent measurement
described in Refs.~\cite{BrEtAl2022}
has led to a value of 
$R_p \approx 0.86 \, {\rm fm}$.
It has very recently been pointed out in Ref.~\cite{Je2022proton}
that two older scattering experiments, carried out
in 1969 at Brookhaven (see Refs.~\cite{CaEtAl1969prl1,CaEtAl1969prl2})
are consistent with an 8\% discrepancy
in the cross sections between
muon-proton and electron-proton scattering,
which translates into 4\% for the
form factor slope, which in turn amounts to 2\% for the radius.
This is precisely the difference
between the ``smaller'' proton radius of
$R_p \approx 0.84 \, {\rm fm}$
and the recently obtained (Ref.~\cite{BrEtAl2022}) value of
$R_p \approx 0.86 \, {\rm fm}$.
The MUSE experiment~\cite{MUSE,KoEtAl2014,Ko2022priv} 
at the Paul--Scherrer Institute aims to remeasure 
the muon-proton cross sections in the near future.

In conclusion, we have shown that 
the idea formulated 
in Refs.~\cite{Lu1988,Ho1998,dV2002,LuEtAl1997}
and Refs.~\cite{JeMoTaWu2008,JeMoTaWu2009,JeMoTa2010}
could lead to a feasible pathway toward a determination
of the Rydberg constant.
This could be interesting because
most recent spectroscopic 
experiments~\cite{BeEtAl2017,FlEtAl2018,BeEtAl2019,BrEtAl2022,YzEtAl2023}
focus on transitions in atomic hydrogen
which depend on both constants in question, namely,
the proton radius and the Rydberg constant.
Focusing on Rydberg states,
as proposed here, means that one
isolates one of these constants, thereby 
potentially obtaining a
clear and distinct picture of the proton radius puzzle.
 The current situation
provides not only motivation to carry out
the MUSE experiment at PSI~\cite{MUSE,KoEtAl2014,Ko2022priv},
but also, to re-double efforts to measure the
Rydberg constant.

\section*{Acknowledgements}

The authors acknowledge extensive insightful 
conversations with B.~J.~Wundt, and helpful 
conversations with S.~M.~Brewer. 
Support from the National Science Foundation
(grant PHY--2110294) is gratefully acknowledged.
Furthermore, U.D.J.~and D.C.Y.~acknowledge
support from the Templeton Foundation (Fundamental Physics Black Grant,
Subawards 60049570 MST and 60049570 CSU 
of Grant ID \#{}61039),
is also gratefully acknowledged.

\end{document}